\documentclass[letterpaper, 10 pt, conference]{ieeeconf}  

\IEEEoverridecommandlockouts                              

\usepackage{cite}
\usepackage{url}
\usepackage{amsmath,amssymb,amsfonts}
\usepackage{multirow}
\usepackage{tabularx}
\usepackage{algorithmic}
\usepackage{graphicx}
\usepackage{hyperref}       
\usepackage{siunitx}
\usepackage{textcomp}
\usepackage{booktabs}

\hypersetup{
    colorlinks=true,
    linkcolor=blue,      
    urlcolor=blue,
    pdftitle={Overleaf Example},
    pdfpagemode=FullScreen,
    }

\usepackage{xcolor}
\def\BibTeX{{\rm B\kern-.05em{\sc i\kern-.025em b}\kern-.08em
    T\kern-.1667em\lower.7ex\hbox{E}\kern-.125emX}}

\begin{document}

\title{\LARGE \bf
Large Car-following Data Based on Lyft level-5 Open Dataset: Following Autonomous Vehicles vs. Human-driven Vehicles
}

\author{Guopeng Li$^{1, *}$,
Yiru Jiao$^1$,
Victor L. Knoop$^1$,
Simeon C. Calvert$^1$, 
and J.W.C. van Lint$^1$
\thanks{\textsuperscript{1} Department of Transport \& Planning, Delft University of Technology, the Netherlands}
}

\maketitle
\thispagestyle{empty}
\pagestyle{empty}

\begin{abstract}

Car-Following (CF), as a fundamental driving behaviour, has significant influences on the safety and efficiency of traffic flow. Investigating how human drivers react differently when following autonomous vs. human-driven vehicles (HV) is thus critical for mixed traffic flow. Research in this field can be expedited with trajectory datasets collected by Autonomous Vehicles (AVs). However, trajectories collected by AVs are noisy and not readily applicable for studying CF behaviour. This paper extracts and enhances two categories of CF data, HV-following-AV (H-A) and HV-following-HV (H-H), from the open Lyft level-5 dataset. First, CF pairs are selected based on specific rules. Next, the quality of raw data is assessed by anomaly analysis. Then, the raw CF data is corrected and enhanced via motion planning, Kalman filtering, and wavelet denoising. As a result, 29k+ H-A and 42k+ H-H car-following segments are obtained, with a total driving distance of 150k+ km. A diversity assessment shows that the processed data cover complete CF regimes for calibrating CF models. This open and ready-to-use dataset provides the opportunity to investigate the CF behaviours of following AVs vs. HVs from real-world data. It can further facilitate studies on exploring the impact of AVs on mixed urban traffic. \\

\end{abstract}


\section{Introduction}\label{Sec:Introduction}

Autonomous vehicles (AVs) have been rapidly developing in recent years, bringing potential benefits such as enhancing traffic safety \cite{Yao2020}, reducing congestion \cite{duarte2018impact}, and increasing mobility accessibility \cite{meyer2017autonomous}. However, the extent of these improvements remains unclear, which depends not only on the performance of AVs but also on human drivers' reactions to AVs. Clarifying the impact of AVs on human driving behaviour is thus crucial for safe and efficient integration of AVs into transportation systems \cite{Hoogendoorn2014, calvert2017will}.

Car-following (CF), which refers to one vehicle following another, is the most common driving behaviour. CF plays a critical role in maintaining smooth traffic flow and reducing congestion \cite{sun2018stability, Shiomi2023}. The presence of AVs may reshape CF behaviours and thus mixed traffic flow. How an AV follows its leading vehicle is determined by the specific driving algorithm, which is continuously being improved. Therefore, it is more important to examine how human-driven vehicles (HVs) react differently when following an AV vs. an HV.

In the literature, the influence of AVs on the CF behaviours of human drivers has mainly been studied through field experiments, driving simulators, and real-world AV datasets. In field experiments, participants are asked to follow a real or seemingly-real AV in different scenarios \cite{Rahmati2019, Mahdinia2021, Soni2022}. Similar experiments can also be carried out in a virtual environment by using driving simulators \cite{reddy2022recognizability}. Field tests and simulations are controllable so researchers can focus on specific points of interest. However, due to cost limitations, these two approaches cannot provide comprehensive and large data covering diverse scenarios.

Recently, the release of autonomous driving datasets, such as Waymo \cite{Sun2020}, nuScenes \cite{Caesar2020}, and Lyft5 \cite{houston2021one}, has enabled researchers to study AVs' impacts on traffic with real-world data. Hu et al. \cite{hu2022processing} offer the first attempt to process a CF dataset from the Waymo dataset. However, because AVs are not marked in the entire dataset of Waymo, only 274 HV-following-AV (H-A) pairs and 1032 HV-following-HV (H-H) pairs are extracted. The limited amount of samples leads to contradictory findings. For example, Wen et al. \cite{wen2022characterizing} conclude that, compared with H-H, H-A has lower driving volatility, smaller time headways, and higher Time-to-Collision (TTC); while Hu et al. \cite{Hu2023} found no significant difference between H-H and H-A, except for smaller spacing during congestion. To reduce the biases when using small datasets, a larger and comparative CF dataset balanced comprising both scenarios is indispensable.

In this paper, the Lyft level-5 open dataset is processed. We select, assess, and enhance 29k+ HV-following-AV pairs and 42k+ HV-following-HV pairs in similar environments. The dataset covers diverse CF regimes and the enhanced dataset provides smooth, ready-to-use motion information for CF model calibration/training. The contribution of this paper is double-fold. First, we propose a processing procedure and demonstrate its validity. Second, the processed data is openly shared as the first large CF dataset that allows comparing the behaviours of following AV vs. HV. It is expected to help better evaluate the influence of AVs in mixed traffic. The dataset is available at \url{https://github.com/RomainLITUD/Car-Following-Dataset-HV-vs-AV}, including detailed instructions on reading and filtering desired CF pairs.

\section{Lyft level-5 dataset}\label{Sec:evaluation}
\subsection{Dataset description}
\begin{figure}[h]
    \centering
    \includegraphics[width=0.9\linewidth]{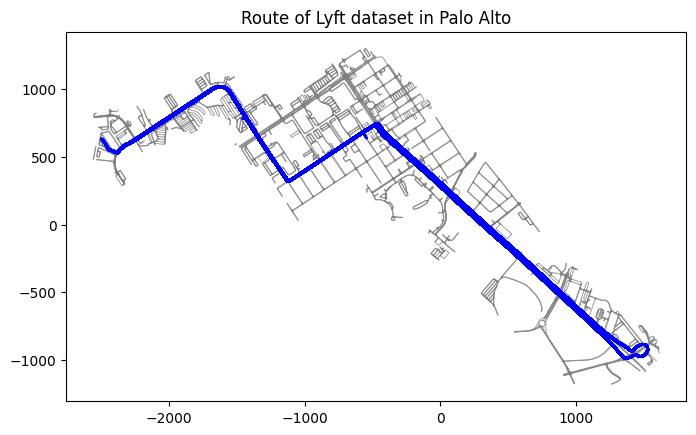}
    \caption{The road maps of Palo Alto, California (in meters). The blue lines mark the fixed route in the Lyft level-5 dataset.}
    \label{fig: route}
\end{figure}

The Lyft level-5 dataset \cite{houston2021one} is a large-scale dataset of high-resolution sensor data collected by a fleet of 20 self-driving cars. The dataset includes 1000+ hours of perception and motion data collected over a 4-month period from urban and suburban environments along a fixed route in Palo Alto, California. The route is shown in Fig.\ref{fig: route}.

The motion prediction dataset comprises about 170,000 scenes, with each scene spanning approximately \SI{25}{\second}. These scenes may be collected continuously or intermittently. For each scene, the track ids of agents are re-numbered (from 1). Each scene includes the movement states (e.g. position, yaw angle, size, speed) of perceived vehicles, cyclists, and pedestrians, as well as the position and orientation of the AV.  Information about the driving environment, including high-definite maps and traffic light status, is also provided. The dataset is available from the website: \url{https://woven.toyota/en/prediction-dataset}. The material used in this paper includes the full training and validation datasets, the semantic map file, and the python toolkit \texttt{l5kit} (\url{https://woven-planet.github.io/l5kit/}) provided by Lyft developers. 

\subsection{Data Processing Framework}

\begin{figure}[h]
    \centering
    \includegraphics[width=0.7\linewidth]{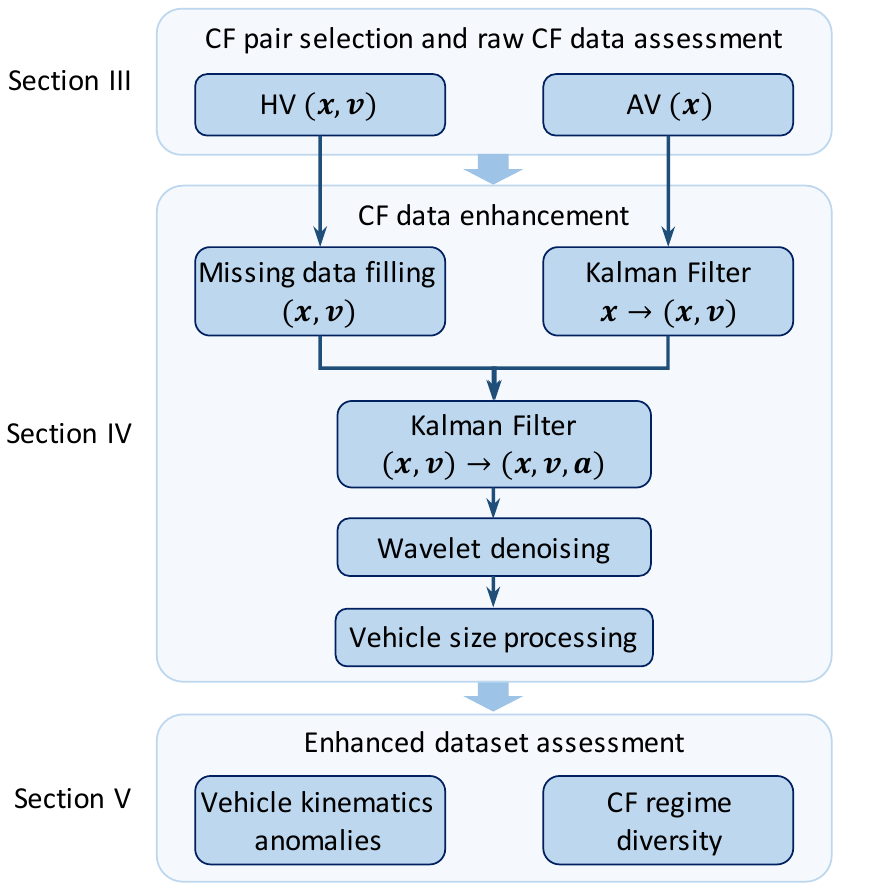}
    \caption{Flowchart of the data processing.}
    \label{fig: flowchart}
\end{figure}

The flowchart in Fig.\ref{fig: flowchart} presents the procedure of CF data selection, assessment, and enhancement. In section \ref{Sec:Methods}, CF pairs and their raw trajectories are first selected from the unlabelled dataset based on certain rules. Next, the raw data quality is assessed from the perspective of anomaly analysis. In section \ref{Sec: enhancement}, the raw data is enhanced. For AVs (only position $x$ is given) and HVs (speed $v$ and position $x$ are given), we use two different methods to fill in the missing segments and estimate smooth $x$ and $v$. The acceleration $a$ is estimated by the combination of the Kalman Filter and wavelet denoising methods. After the processing procedure, in section \ref{Sec:Discussion}, the enhanced data is assessed again, including both anomaly analysis and the diversity of CF regimes. In the following sections, we will introduce each step in detail.

\section{CF pair selection and assessment}\label{Sec:Methods}

\subsection{CF pair selection}

The first step is identifying CF pairs from unlabelled data. Previous studies about the rules of extracting CF pairs have been reviewed by Wen et al. \cite{wen2022characterizing}. We refer the readers to this recent paper for more details.

Considering the large size of the dataset (1000+ hours), we make two groups of rules and propose a two-step procedure. The first step is quick screening. The entire dataset is scanned second by second (every 1-5 seconds) based on the first group of rules to roughly identify possible CF events. Next, the second group of rules rigorously check each CF event frame by frame. The rules are listed in Table.\ref{tab: CF selection rules}. The extracted CF pairs are categorized based on the type of the leading vehicle (AV or HV). Statistics of the CF dataset are presented in Table.\ref{tab: CF statistics}. The total duration of this CF dataset spans over 460+ hours, covering a total distance of 15,000+ km.

\begin{table}[h]
\centering
\caption{CF event selection rules \cite{wen2022characterizing} \label{tab: CF selection rules}}
\begin{tabular}{l p{0.85\linewidth}}
\toprule
No. & Description \\
\midrule
\multicolumn{2}{c}{Group-1: second-to-second}\\
\midrule
1.1 & The probability of both vehicles being passenger cars $>$ 0.95.\\
1.2 & The longitudinal distance between the vehicles $<$ \SI{85}{\meter}.\\
1.3 & The lateral distance between the vehicles $<$ \SI{1.75}{\meter}.\\
1.4 & No obstacles or other road agents interrupt the CF pairs.\\
1.5 & Both vehicles are on the same side of the traffic lights.\\
1.6 & Both vehicles are on a straight road.\\
1.7 & Rules 1.1-1.6 must hold with a duration $>$ \SI{16}{\second}.\\
\midrule
\multicolumn{2}{c}{Group-2: frame-to-frame}\\
\midrule
2.1 & The deviation of yaw for both vehicles $<$ 0.035 rad.\\
2.2 & The maximum yaw to the driving direction for both vehicles $<$ 0.087 rad.\\
2.3 & The interval between all adjacent timestamps $<$ \SI{0.42}{\second}.\\
2.4 & The distance between all adjacent positions $<$ \SI{5}{\meter}.\\
2.5 & The average speed of both vehicles $>$ \SI{1}{\meter\per\second}.\\
\bottomrule
\end{tabular}
\end{table}

\begin{table}[h]
    \caption{Extracted CF pairs from Lyft dataset}
    \label{tab: CF statistics}
    \begin{center}
    \vspace{-10pt}
    \begin{tabular}{l|ccc}
    \toprule
    \textbf{Dataset} & \textbf{CF pairs}  & \textbf{Distance(km)} & \textbf{Duration(h)}\\ 
    \midrule
    H-A & 29449 & 6996 & 220\\
    H-H & 42892 & 8125 & 240\\
    \bottomrule
    \end{tabular}
    \end{center}
\end{table}
For each CF pair, the initial position of the leading vehicle is set as the origin. The straight road lane defines the driving direction. Because the angle between the yaw of the vehicles and the lanes is small (as stated in Rules 2.1 and 2.2), it is reasonable to assume that the lateral movement is independent of the car-following behaviours. Therefore, this paper will only focus on the longitudinal movement going forward. 

\subsection{Raw CF data assessment}\label{sec: raw eva}

Calibrating CF models requires highly accurate and consistent position, speed, and acceleration data. It is necessary to first assess the quality of the raw CF data. For HV (either in H-A or H-H dataset), we assess (1) anomalies that violate constraints of vehicle kinematics, and (2) missing data. For AV, because only $x$ is given, the anomalies of vehicle kinematics are assessed only. According to Punzo et al. \cite{punzo2011assessment}, 3 constraints on acceleration $a$ and jerk $j$ must be satisfied:
\begin{itemize}
    \item $a\in [-8, 5]\ \SI[per-mode=repeated-symbol]{}{\meter\per\second\squared}\label{eq: range}$
    \item $j\in [-15, 15]\ \SI[per-mode=repeated-symbol]{}{\meter\per\cubic\second}$
    \item The jerk's sign cannot be inverse more than once in \SI{1}{\second} (denoted as \textit{Jerk Sign Inversion}, JSI).
\end{itemize}
For HV, $a$ and $j$ are directly computed by the 1st and the 2nd-order differentiation of $v$ ($v$-based). While for AV, $a$ and $j$ are derived from the 2nd and the 3rd-order differentiation of $x$ ($x$-based). Violating one of the 3 constraints is identified as an anomaly. The result is shown in Table.\ref{tab: consistency}, which is also compared with the CF data in the Waymo dataset (evaluated in \cite{hu2022processing}). The $v$-based jerk and JSI anomaly proportion for HVs in Lyft are on-par with the Waymo dataset. For $x$-based AVs, Lyft shows higher quality than Waymo (lower anomaly proportion).

\begin{table}[h]
    \caption{Anomaly assessment of raw CF data}
    \label{tab: consistency}
    \begin{center}
    \vspace{-10pt}
    \begin{tabular}{l|cc|cc}
    \toprule
     & \multicolumn{2}{c}{Lyft} & \multicolumn{2}{c}{Waymo}\\
     \midrule
    \textbf{Statistics} & \shortstack{AV\\($x$-based)} & \shortstack{HV\\($v$-based)} & $x$-based & $v$-based\\ 
    \midrule
    Anomaly acc (\%) & 0.004 & 0.226 & -- & --\\
    Anomaly jerk (\%) & 1.007 & 0.656 & 5.3 & 0.439\\
    Anomaly JSI (\%) & 34.2 & 25.4 & 86.1 & 37.2\\
    \bottomrule
    \end{tabular}
    \end{center}
\end{table}

\begin{figure}[h]
    \centering
    \includegraphics[width=\linewidth]{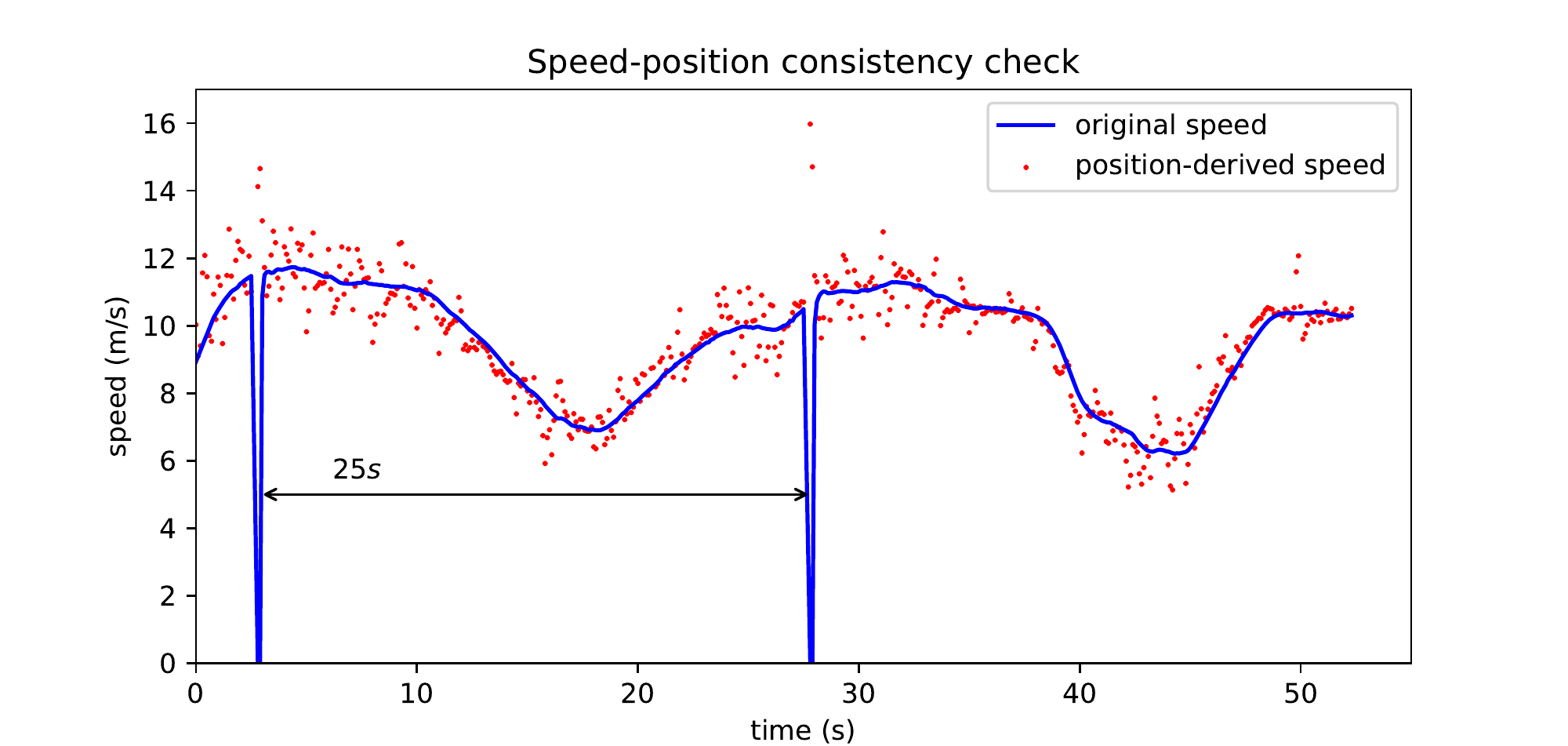}
    \caption{Comparison between the speed data and the position-derived speed.}
    \label{fig: consistency}
\end{figure}

Next, abnormal segments are further investigated. Fig.\ref{fig: consistency} shows an example that compares the given speed and $x$-based speed of an HV. The result demonstrates that the speed provided by Lyft is smoothed from the measurements of position. However, there also exist some errors. At the beginning and the end of each scene (\SI{25}{\second} duration), there is a 0-speed value, which seems to have been added artificially and was not excluded in smoothing. We guess that perhaps these 0-values are used to assist in segmenting scenes (for deep-learning purposes) but for some reason not removed. Therefore, the first 0.5 to \SI{1.5}{\second} data in these scenes are unreliable. 

The assessment of the raw CF data is summarized as follows:
\begin{itemize}
    \item The given position data of AVs is of high quality.
    \item Data are missing at the beginning and end of each scene.
    \item For both AVs and HVs, acceleration and jerk derived by differentiation is not smooth enough for calibrating CF models.
\end{itemize}

In the next section, the raw data will be enhanced to address these problems.

\section{CF Data enhancement}\label{Sec: enhancement}
\subsection{Missing data filling}

Before further processing raw HV data, the missing part around the 0-speed points must be filled in first. In this study, we remove the speed and position data within \SI{1.5}{\second} of the last 0-value timestamp. This segment of motion is estimated by the polynomial of degree seven (7-DOP) jerk-minimization method. The position of the vehicle is assumed to have the form of a polynomial of degree seven with 8 unknown coefficients ${p_i}_{i=0}^7$:
\begin{equation}
    x(t) = \sum_{i=0}^7 p_i t^i
\end{equation}

The initial and final positions, speeds, and accelerations (derived from speed) pose 6 boundary conditions (constraints). The objective function to minimize is the jerk in the duration $T$:
\begin{equation}
    J = \int_{0}^T [x'''(t)]^2 dt
\end{equation}
which is typical quadratic programming that can be easily solved. Notice that only the position and speed are estimated. The acceleration data will be derived later.

\begin{figure}[h]
    \centering
    \includegraphics[width=\linewidth]{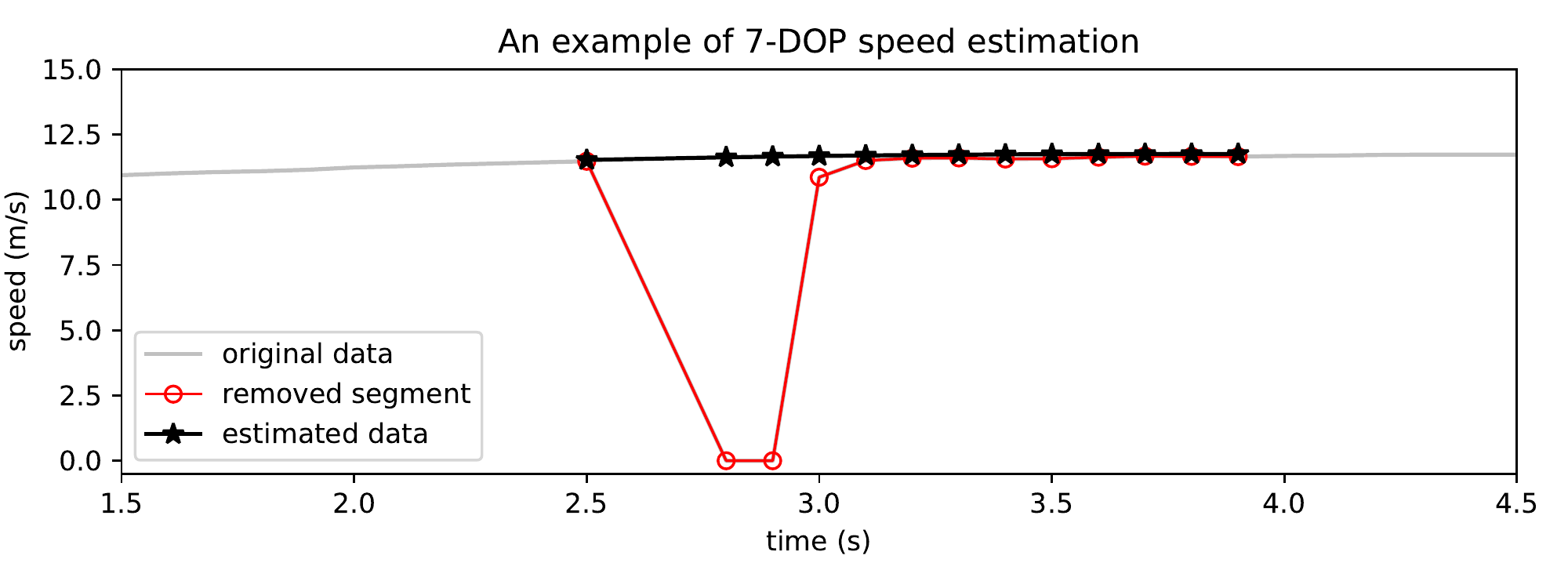}
    \caption{Use the 7-DOP method to fill in the incorrect data segment.}
    \label{fig: 7dop}
\end{figure}

One example of the estimation results is shown in Fig.\ref{fig: 7dop}. The time interval is not always uniform. Clearly, the 0-value influences the given speed at around \SI{3.0}{\second} (the red line). The removed segment is filled in by the black-star curves (the timestamp series are not changed), which is a 6-degree polynomial. We see that the estimated speed profile is smoother due to the vehicle's kinematic constraints.

\subsection{Kalman filtering for speed estimation}

Because the speed of AVs is not provided, to facilitate the following acceleration estimation and smoothing steps, we need to derive the speed of AVs from position data first. The well-known Kalman Filter (KF) \cite{welch1995introduction} is used. In this Kalman Filter, we employ the constant-speed model. For each time interval (mostly $\Delta t = \SI{0.1}{\second}$), the state transition equation is:

\begin{equation}
    \begin{bmatrix}
    x(t_{i+1})\\
    v(t_{i+1})
    \end{bmatrix}
    =
    \begin{bmatrix}
    1 & \Delta t_i\\
    0 & 1
    \end{bmatrix}
    \cdot
    \begin{bmatrix}
    x(t_{i})\\
    v(t_{i})
    \end{bmatrix}
\end{equation}
The given position and its differentiation are regarded as the measurement of $x$ and $v$. The process covariance matrix $\mathbf{Q}_1$ and the measurement covariance matrix $\mathbf{R}_1$ control the trade-off between accuracy and smoothness. Considering the range of acceleration in Eq.\ref{eq: range}, their values are set as follows by trial and error:
\begin{equation}
    \mathbf{Q}_1
    =
    \begin{bmatrix}
    0.2 & 0\\
    0 & 0.8
    \end{bmatrix}^2,\ \ 
    \mathbf{R}_1
    =
    \begin{bmatrix}
    0.5 & 0\\
    0 & 1.1
    \end{bmatrix}^2
\end{equation}

The next step is estimating the acceleration. In principle, this can also be accomplished by a Kalman filter that gives both speed and acceleration. However, tuning the process and measurement covariance is difficult and usually the acceleration covariance changes with time for a CF event. Therefore, we propose a 2-step method. First, continuing in this subsection, a second KF is applied for both AVs and HVs to derive an over-smoothed acceleration. Next in subsection \ref{Sec:WaveletDenoise}, wavelet denoising will be used to smooth the $v$-based (differentiation) acceleration. For the second KF, the state transition equation is:
\begin{equation}
    \begin{bmatrix}
    x(t_{i+1})\\
    v(t_{i+1})\\
    a(t_{i+1})
    \end{bmatrix}
    =
    \begin{bmatrix}
    1 & \Delta t_i & \frac{1}{2}a(t_{i})(\Delta t_i)^2\\
    0 & 1 & a(t_{i})\Delta t_i\\
    0 & 0 & 1
    \end{bmatrix}
    \cdot
    \begin{bmatrix}
    x(t_{i})\\
    v(t_{i})\\
    a(t_{i})
    \end{bmatrix}
\end{equation}
The $\mathbf{Q}_2$ and $\mathbf{R}_2$ are set as follows. We set a high value of measurement error for acceleration to over-smooth it.
\begin{equation}
    \mathbf{Q}_2
    =
    \begin{bmatrix}
    0.2 & 0 & 0\\
    0 & 0.4 & 0\\
    0 & 0 & 1.5
    \end{bmatrix}^2,\ \ 
    \mathbf{R}_2
    =
    \begin{bmatrix}
    0.5 & 0 & 0\\
    0 & 1 & 0\\
    0 & 0 & 10
    \end{bmatrix}^2
\end{equation}

Denote the $v$-based acceleration as $a_v$ and the over-smoothed acceleration given by KF as $a_k$, their RMSE, $\sigma_a$, is regarded as an approximation of the total noise. This estimated hyper-parameter will be used in wavelet denoising to further smooth acceleration.

\subsection{Wavelet denoising for acceleration smoothing}\label{Sec:WaveletDenoise}

Wavelet denoising \cite{pan1999two} is a robust technique to remove noise from signals by transforming the time series into wavelet space and then thresholding the high-frequency coefficients while preserving the important information in the low-frequency coefficients. Compared with KF, wavelet denoising is non-parametric. It is effective for removing different types of noise, including Gaussian noise, impulsive noise, and mixed noise. Especially, wavelet denoising can adapt to non-stationary noise characteristics, which meets the requirements of acceleration estimation and smoothing.

In this study, we use the wavelet denoising tool in \texttt{skimage} python package. The noise standard deviation is set as $\sigma_a$ from the result of KF. The type of wavelet is the Daubechies family with 6 vanishing moments (`db6'), which can effectively capture both short-term and long-term features. A soft threshold method and up to 4 wavelet decomposition levels are used.

\begin{figure}[h]
    \centering
    \includegraphics[width=\linewidth]{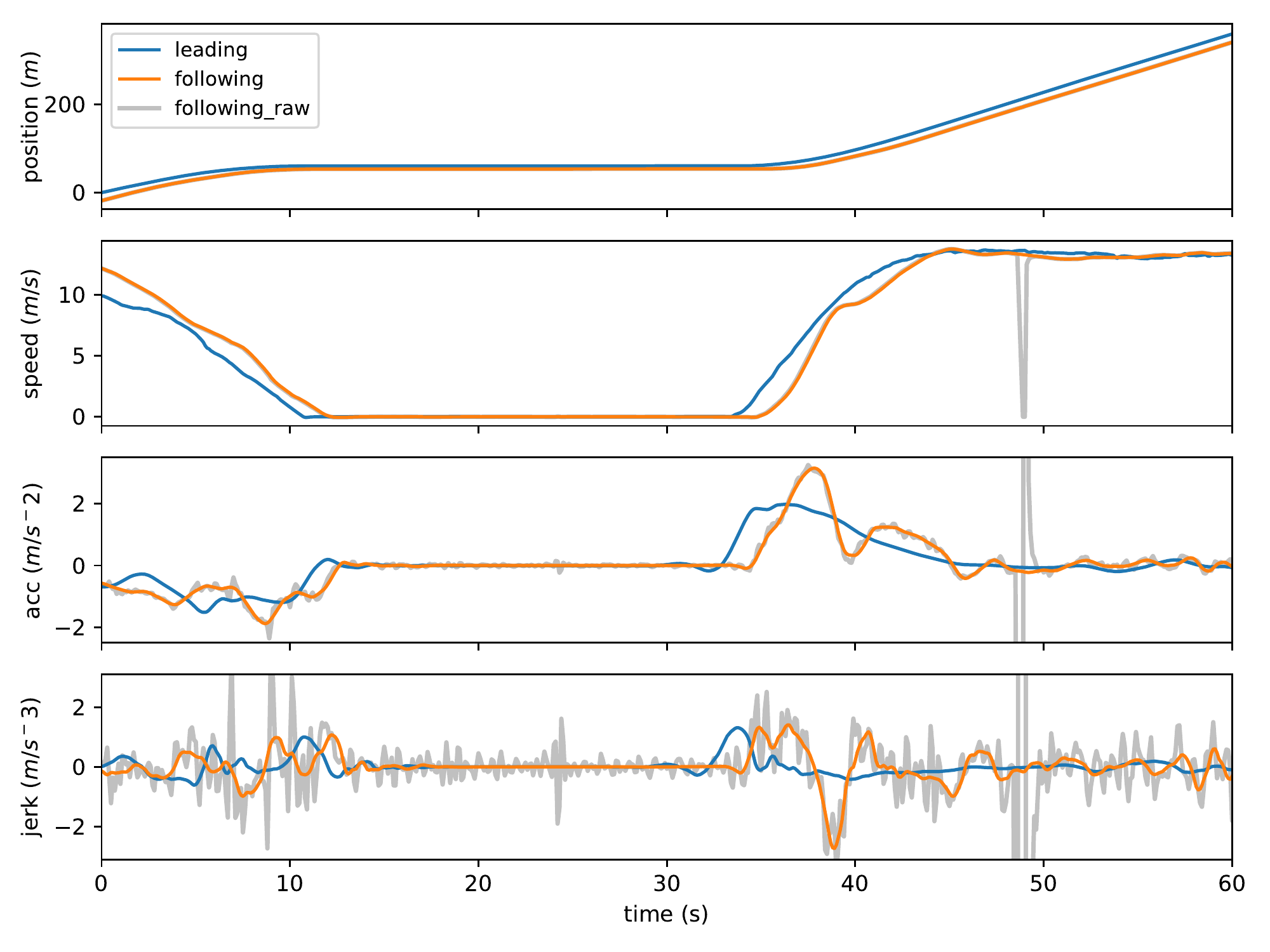}
    \caption{Comparison between the raw data and the processed data in a \SI{60}{\second} CF event.}
    \label{fig: example}
\end{figure}

Fig.\ref{fig: example} presents a \SI{1}{\minute} H-A car-following example. The silver lines are raw position, speed, $v$-based acceleration and jerk of the following HV. This CF pair approaches an intersection, decelerates, stops and waits, and then accelerates to a desired speed. The missing segments at \SI{24}{\second} and at \SI{49}{\second} are filled in. Processed acceleration and jerk profiles are significantly smoother.

\subsection{Vehicle size processing}

Besides motion information, the length of the vehicle is also important, especially for calculating some safety and efficiency metrics, such as bump-to-rear gap and TTC. The size of the AVs is a given, fixed value (\SI{4.87}{\meter} length, \SI{1.85}{\meter} width). However, the size of perceived HVs is not always stable. It varies with time due to perception errors (such as shading). Many datasets, including the Waymo CF data \cite{hu2022processing}, ignore this step.

\begin{figure}[h]
    \centering
    \includegraphics[width=0.75\linewidth]{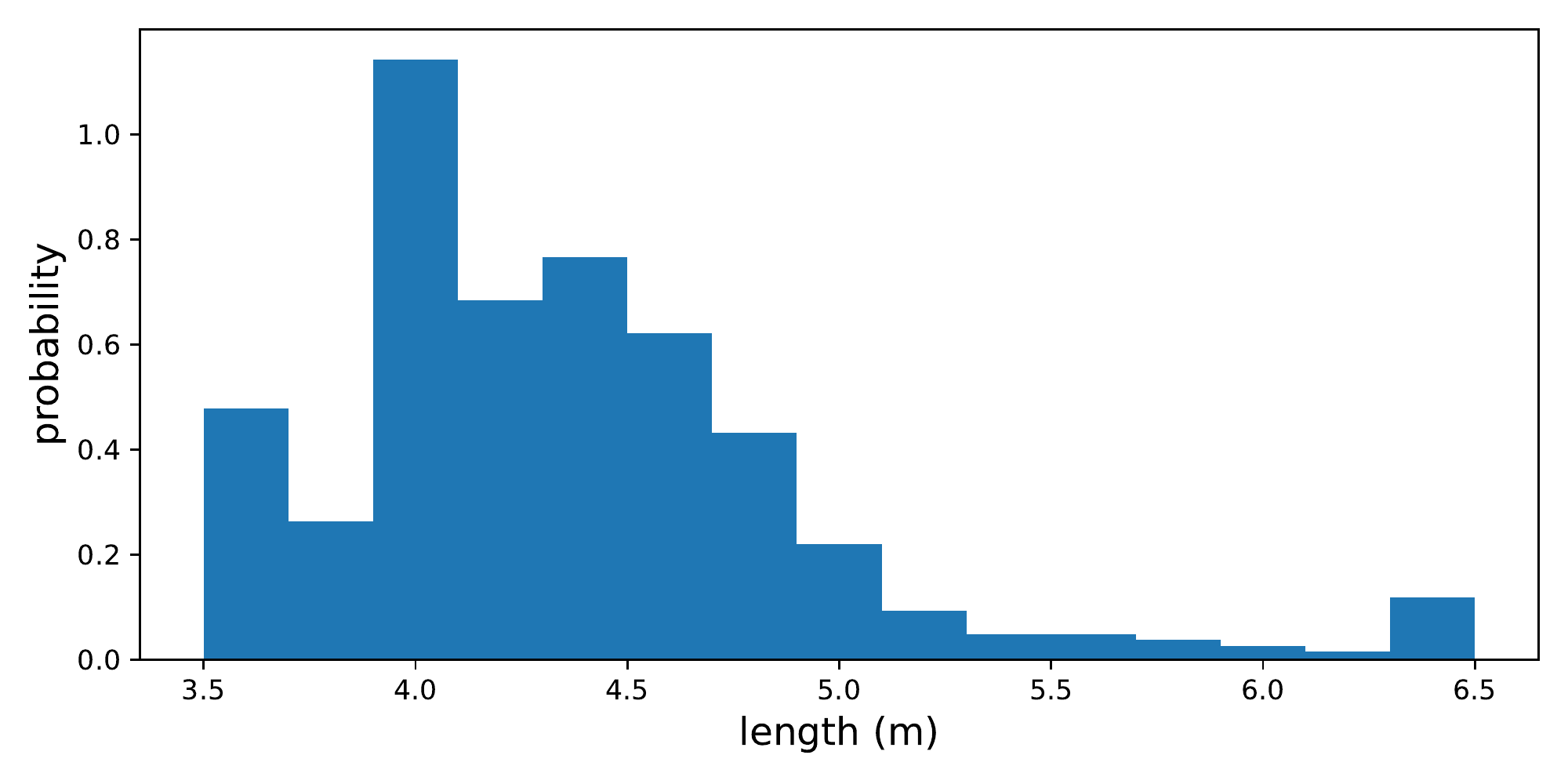}
    \caption{Distribution of HV lengths.}
    \label{fig: length}
\end{figure}

Because we excluded vehicles that are not passenger cars when selecting CF pairs, we set a maximum length of \SI{6.5}{\meter} (pick-up truck) and a minimum length of \SI{4}{\meter} (small vehicle). The following rules are used to estimate the size. (1) For each HV, if the variance of the length time series $<$ \SI{0.3}{\meter}, then we choose the mean value. (2) Otherwise, perceived length series are clamped between \SI{3.5}{\meter} and \SI{6.5}{\meter}, then we choose the percentile at 0.95 (95\%). The distribution of estimated HV lengths is shown in Fig.\ref{fig: length}. The mean value is \SI{4.38}{\meter}, which is close to the average length of passenger cars in the US \cite{tiwari2000passenger}.

By now, the entire data enhancement procedure is finished. Next, we will assess the quality of enhanced data.

%

\section{Enhanced dataset evaluation}\label{Sec:Discussion}

The enhanced dataset is split into two groups, following AV (denoted as H-A) vs. following HV (denoted as H-H). They will be assessed based on anomaly analysis and diversity evaluation. We will show that the enhanced trajectories have higher quantity and the dataset covers diverse regimes for calibrating CF models.

\subsection{Vehicle kinematics anomalies}

For anomaly detection, we use the same rules as stated in section \ref{sec: raw eva}. The result is shown in Table.\ref{tab: assess-enhanced}. After processing, the anomaly percentage is significantly reduced compared to the raw data in Table.\ref{tab: consistency}, especially abnormal jerk sign inversion. The enhanced data is smoother and better conforms to the vehicle's kinematic constraints.

\begin{table}[h]
    \caption{Anomaly assessment of the enhanced dataset}
    \label{tab: assess-enhanced}
    \begin{center}
    \vspace{-10pt}
    \begin{tabular}{l|ccc}
    \toprule
    \textbf{Statistics} & H-A pairs & H-H pairs\\ 
    \midrule
    Anomaly acc (\%) & 0.0082 & 0.0234\\
    Anomaly jerk (\%) & 0.0039 & 0.0186\\
    Anomaly jerk sign inversion (\%) & 0.455 & 0.454\\
    \bottomrule
    \end{tabular}
    \end{center}
\end{table}

\subsection{CF regime diversity}

Next, we will evaluate the regime diversity in both H-H and H-A subsets. Here a regime refers to a driving situation experienced by the following vehicle (usually restricted by its leader). A higher diversity of regimes means that the dataset is not just `big' but also informative. For some CF models, e.g. Intelligent Driver Models (IDM), certain parameters cannot be calibrated if some specific regimes are missing in the data \cite{Treiber2013}. Insufficient regime diversity in CF data may also result in unrepresentative or over-fitted models \cite{Sharma2019}, and thus contradictory conclusions about driving behaviours. 

\begin{figure}[h]
    \centering
    \includegraphics[width=0.75\linewidth]{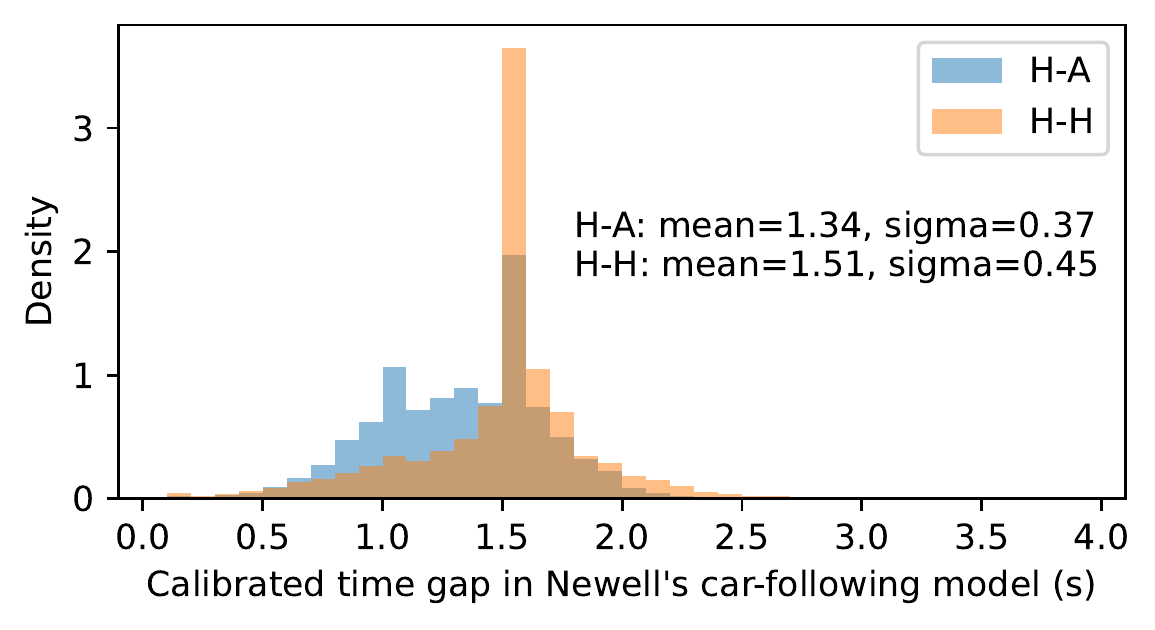}
    \caption{Distribution comparison between H-H and H-A time gaps calibrated from Newell's car-following model.}
    \label{fig: tau_estimation}
\end{figure}

To evaluate the regime diversity in the enhanced dataset, we adopt the identification algorithm proposed in \cite{Sharma2018}. The algorithm consists of three steps: 1) segmenting the follower's speed profile into various sections, 2) categorizing the sections into car-following (CF) and free-following (FF), and 3) determining regimes based on the acceleration within these sections. For the second step, a threshold is selected based on the mean and variance of the distribution of time gaps that are calibrated from Newell's car-following model \cite{Newell2002} (we refer the readers to \cite{Sharma2018} for more details). Given that the followers in H-H and H-A may exhibit different time gap distributions, we perform two separate threshold selection. Fig.\ref{fig: tau_estimation} compares the distributions of time gaps for H-H and H-A, where a clear difference is shown between following an HV vs. an AV.

Sharma et al. \cite{Sharma2018} classify regimes based on the previous studies by Treiber et al. \cite{Treiber2013}, where standstill and deceleration are excluded from free-flowing. This consideration is reasonable for highways because the traffic flow on highways is highly continuous. However, vehicles face more interruptions in urban environments, such as vulnerable road agents (cyclists, pedestrians) and traffic signals. Therefore, this paper considers the following 7 regimes: free acceleration (\textbf{Fa}), free deceleration not caused by the leading vehicle (\textbf{Fd}), cruising at a desired speed (\textbf{C}), acceleration following a leading vehicle (\textbf{A}), deceleration following a leading vehicle (\textbf{D}), constant speed following (\textbf{F}), and standstill (\textbf{S}). 

\begin{figure}[h]
    \centering
    \includegraphics[width=0.8\linewidth]{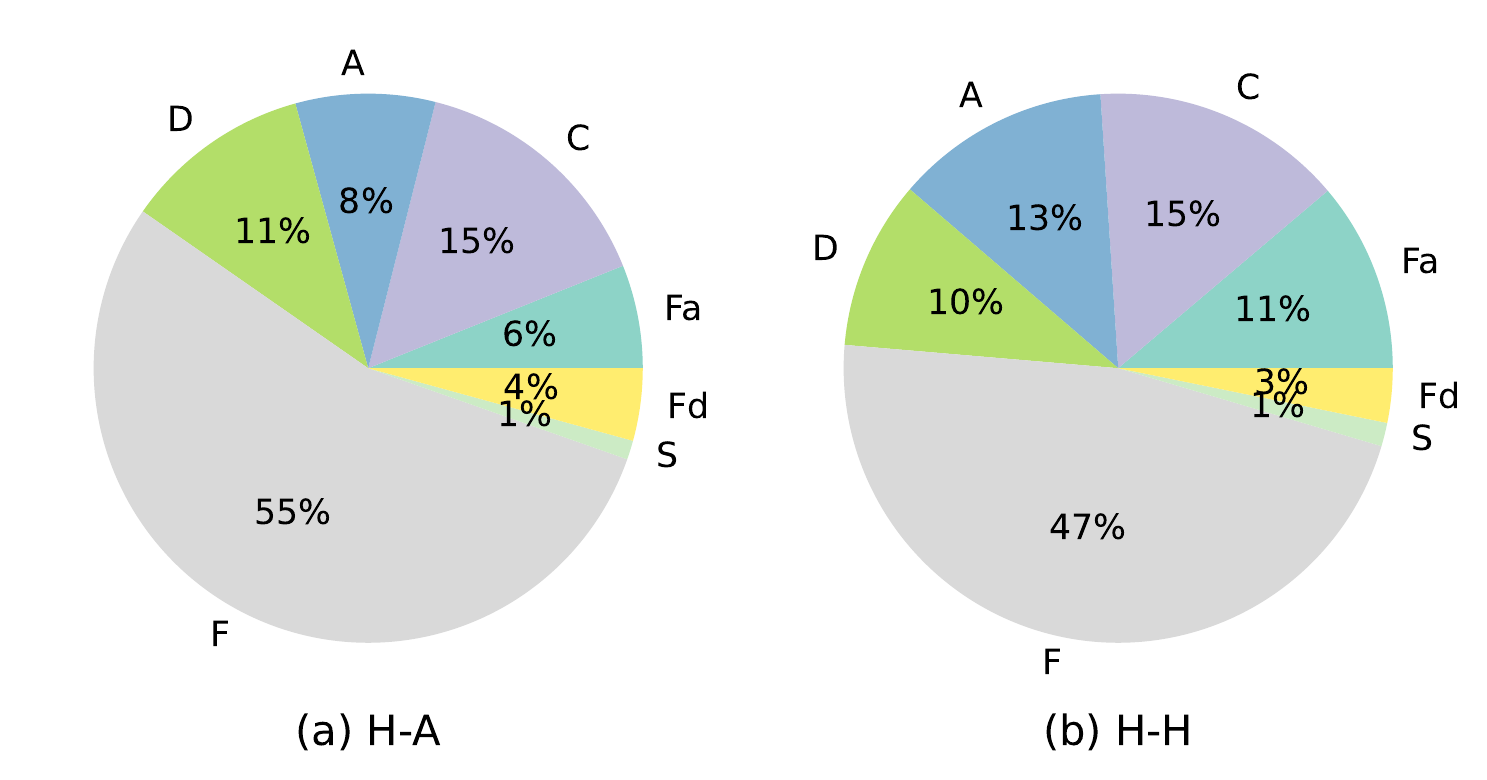}
    \caption{Time proportion of car-following regimes.}
    \label{fig: regime_time}
\end{figure}

Fig.\ref{fig: regime_time} presents the proportion of accumulated duration of the 7 regimes in H-A and H-H subsets. The regimes F, D, and A constitute around 74\% and 70\% of the total duration in H-A and H-H, respectively. 
This suggests that the followers' behaviours depend on their leaders most of the time. This dataset is therefore suitable for studying how followers react differently to the leading AV vs. HV. 

\begin{figure}[h]
    \centering
    \includegraphics[width=0.85\linewidth]{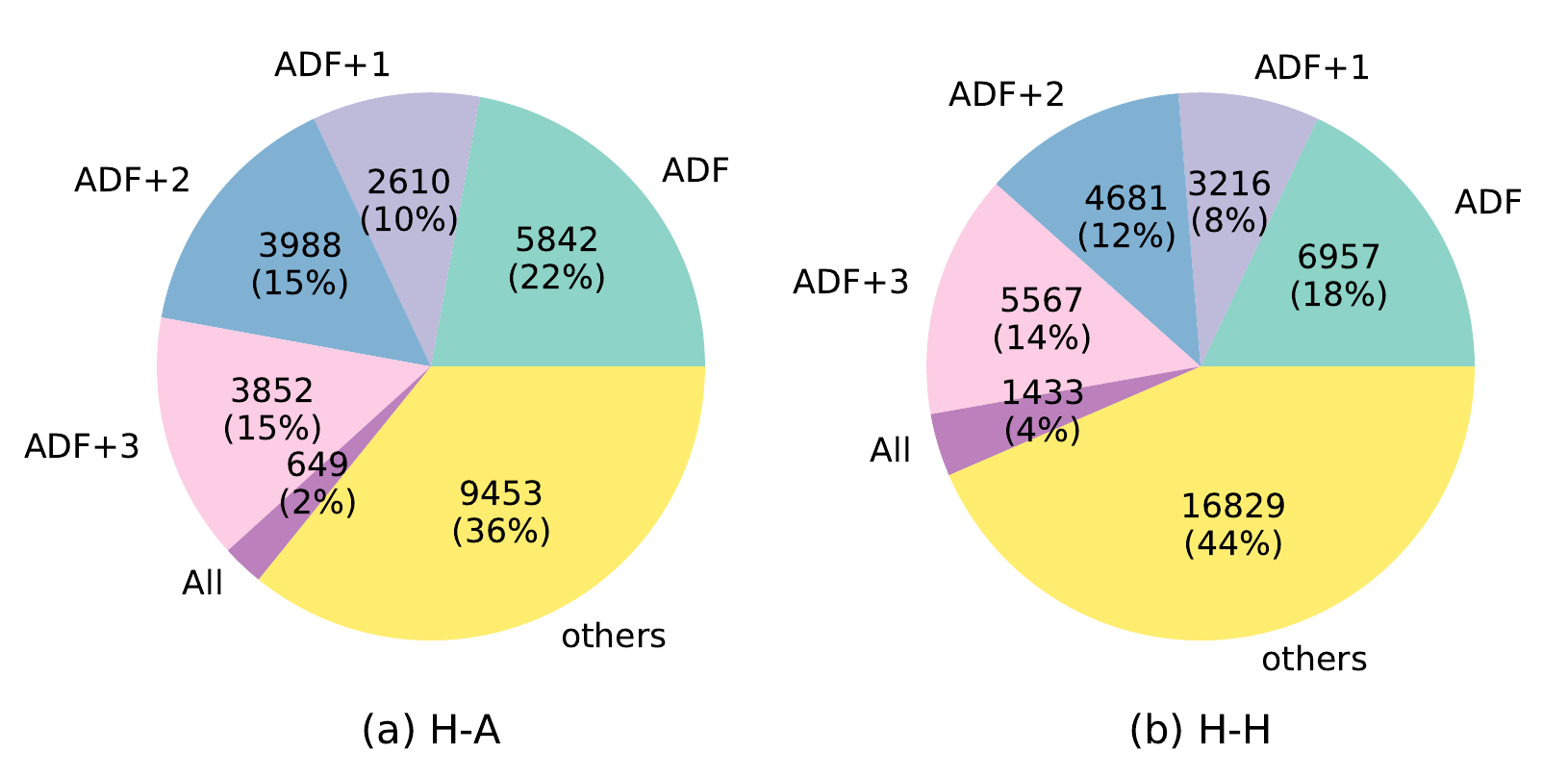}
    \caption{Regime diversity of car-following pairs.}
    \label{fig: regime_pairs}
\end{figure}

According to Sharma et al. \cite{Sharma2019}, to calibrate all parameters in IDM, at least 3 regimes, A, D, and F, must be included in a CF event. Based on this principle, we categorize all CF pairs into two groups. One is `ADF+$n$', which means all ADF regimes are included and there are $n$ extra regimes. The other is `others', which means at least one of ADF regimes is missing (no matter how many regimes they have). The results in Fig.\ref{fig: regime_pairs} show that 64\% and 56\%  CF pairs fall in the ADF+$n$ group in H-A and H-H subset, respectively. These pairs support calibrating IDMs. Meanwhile, we would like to emphasize that `others' do not mean that these CF pairs are useless. They are not suitable for calibrating IDM but they are still informative for calibrating other more complex car-following models, e.g. deep-neural-networks-based methods.

In summary, this section shows that the enhanced dataset has fewer anomalies and high car-following regime diversity.

\section{Conclusion}\label{Sec:Conclusion}

This paper proposes a car-following trajectory data processing procedure. This procedure has been applied to an openly available dataset and validated by anomaly analysis and regime assessment. The Lyft level-5 dataset, which contains information about both autonomous vehicles and human-driven vehicles, has been processed with this technique and the enhanced car-following trajectories are publicly available. The initial dataset now is processed into a high-quality, ready-to-use dataset. It contains human drivers following autonomous as well as human-driven vehicles in diverse scenarios. The processing procedure is essential for doing further analysis. The published enhanced car-following dataset is expected to help researchers better understand the impact of AVs on traffic flow and to develop safer and more effective AV systems in the future.



\addtolength{\textheight}{-12cm}   

\section*{ACKNOWLEDGMENT}

This research is sponsored by the NWO/TTW project MiRRORS (16270) and the TU Delft AI Labs programme.


\bibliographystyle{ieeetr}
\bibliography{references}

\end{document}